\documentclass{elsart}

\usepackage{graphicx}
\usepackage{subfigure}

\def\ca{\c{c}\~{a}}

\begin{document}

\begin{frontmatter}

\title{Quadrupole polarizabilities of the pion in the Nambu--Jona-Lasinio 
model\thanksref{grant}} 

\thanks[grant]{Research supported by the Polish Ministry of Science and Higher
Education, grants N202~034~32/0918 and N202~249235, by Funda\ca o para a 
Ci\^encia e Tecnologia, grants FEDER, OE, POCI 2010, CERN/FP/83510/2008, and 
by the European Community-Research 
Infrastructure Integrating Activity "Study of Strongly Interacting Matter" 
(Grant Agreement 227431) under the Seventh Framework Programme of EU.} 

\author{B. Hiller$^1$, W. Broniowski$^{2,3}$, A. A. Osipov$^{1,4}$, A. H. 
Blin$^1$}

\address{$^1$Centro de F\'{\i}sica Computacional, Departamento de 
F\'{\i}sica da Universidade de Coimbra, 3004-516 Coimbra, Portugal}
\address{$^2$The Niewodnicza\'nski Institute of Nuclear Physics, Polish 
Academy of Sciences, PL-31342 Cracow, Poland}
\address{$^3$Institute of Physics, \'Swi\c{e}tokrzyska Academy, 
PL-25406~Kielce, Poland}
\address{$^4$Dzhelepov Laboratory of Nuclear Problems, JINR 141980 Dubna, 
Russia}

\begin{abstract}
The electromagnetic dipole and quadrupole polarizabilities of the neutral 
and charged pions are calculated in the Nambu--Jona-Lasinio model. Our 
results agree with the recent experimental analysis of these quantities 
based on Dispersion Sum Rules. Comparison is made with the results from  
the Chiral Perturbation Theory.

\end{abstract}
\end{frontmatter}

 
The obtainment of the pion structure parameters, in particular its 
electric and magnetic polarizabilities \cite{Klein:1955}, has challenged 
the experimental and theoretical community since the early sixties 
\cite{Pomeranchuk:1961,Petrun'kin:1964} and today remains a field 
of intense activity (see e.g. \cite{Fil'kov:2008}). Indeed, on the 
experimental side we expect in the future more precise measurements of 
the pion polarizabilities by the COMPASS collaboration at CERN 
\cite{Abbon:2007}, which probably will help to resolve the 
discrepancy between the dispersion sum rule (DSR) and the chiral 
perturbation theory ($\chi$PT) for the difference of the dipole 
polarizabilities $(\alpha_1 -\beta_1 )_{\pi^\pm}$\cite{Pasquini:2008ep,talks}. 
On the theoretical side there is also growing interest in the determination 
of the higher order structure characteristics -- the quadrupole 
polarizabilities of the pion. It has been reported that the DSR values of 
the quadrupole polarizabilities $(\alpha_2\pm\beta_2)_{\pi^\pm}$ and 
$(\alpha_2+\beta_2)_{\pi^0}$ \cite{Fil'kov:2005,Fil'kov:2006,Fil'kov:2007} 
disagree with the present two-loop $\chi$PT calculations 
\cite{Gasser:2005, Gasser:2006}.  

We contribute to this study by calculating the pion electromagnetic 
polarizabilities within the Nambu--Jona-Lasinio model (NJL) \cite{Nambu:1961}. 
To our knowledge, this is the first dynamical calculation for the 
{\em quadrupole} polarizabilities. The NJL model takes into account the 
quark-antiquark structure of the pion explicitly, providing an example of 
dynamical chiral symmetry breaking in a system with nonlinear four-quark 
interactions. In Ref.~\cite{Bajc:1996} the lowest order (dipole) 
polarizabilities have been computed. The alternative approach within the 
large-distance expansion of the extended NJL model has been used in 
\cite{Osipov:1985a}, and for $\gamma\gamma\to\pi^0\pi^0$ mode in 
\cite{Bellucci:1995,Bel'kov:1996,Bijnens:1996}. Here we extend the study 
of Ref.~\cite{Bajc:1996} to the quadrupole case, and find a very reasonable 
agreement with the experimental data. For instance, with the empirical 
values $m_{\pi^0}=136$, $f_\pi=93.1$~MeV, and with the quark mass $M=300$~MeV, 
we obtain the value $(\alpha_2+\beta_2)_{\pi^0}=-0.144\cdot 10^{-4}$ fm$^{5}$, 
which agrees within the error bars with the value from DSR, 
$(\alpha_2+\beta_2)_{\pi^0}=-0.171\pm 0.067 \cdot 10^{-4}$ fm$^{5}$ 
\cite{Fil'kov:2005}. This is an example of the general behavior found here: 
the results of the NJL calculations of the dipole and quadrupole 
polarizabilities are in a good agreement with the DSR values.

\begin{figure}[tb]
\begin{center}
\subfigure{\includegraphics[width=0.28\textwidth]{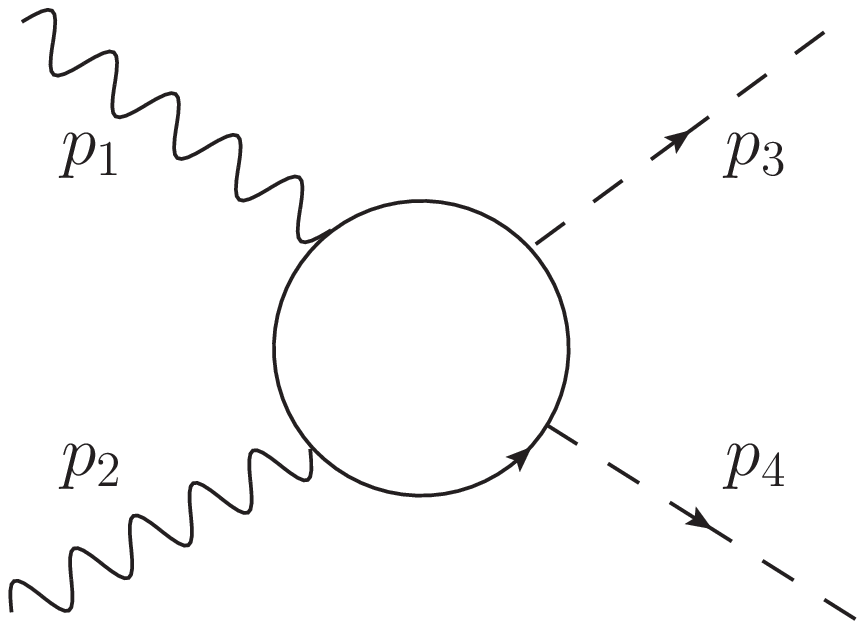}}
\subfigure{\includegraphics[width=0.35\textwidth]{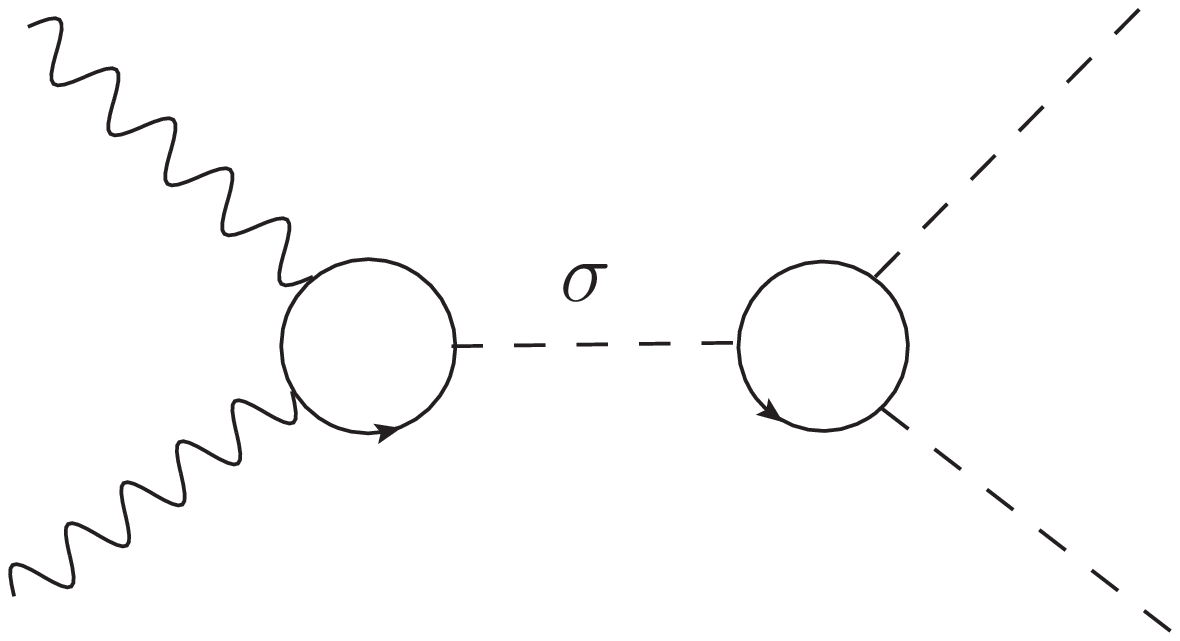}}
\subfigure{\includegraphics[width=0.29\textwidth]{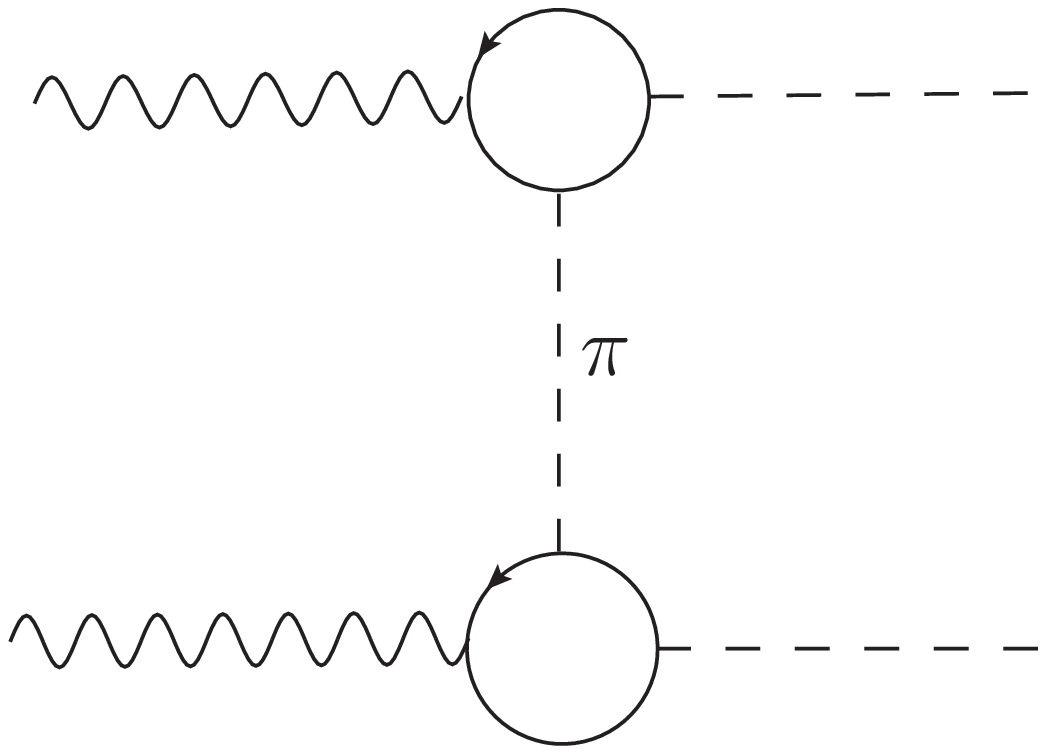}}
\end{center}
\caption{Leading-$N_c$ quark-loop diagrams for the $\gamma\gamma\to\pi\pi$ 
amplitude. The crossed terms are not displayed. 
\label{fig:graphs}}
\end{figure}

Our calculations at the leading-$N_c$ order are done according to the Feynman 
diagrams of Fig.~\ref{fig:graphs}. The analytic expressions for the $A$ and 
$B$ amplitudes, defined below, are derived with the method presented in detail 
in Ref.~\cite{Bajc:1996}. It is also necessary to include the 
$1/N_c$-suppressed one-pion-loop contribution, as discussed below. The 
amplitudes are functions of the Mandelstam variables related to the 
$\gamma(p_1,\epsilon_1)+\gamma(p_2,\epsilon_2)\to\pi^a(p_3)+\pi^b(p_4)$ 
reaction with the on-shell pions and photons. The amplitude  
\begin{equation}
\label{amplitudes}
   T(p_1,p_2,p_3)=e^2\epsilon_1^\mu\epsilon_2^\nu T_{\mu\nu}, \quad
   T_{\mu\nu}=A(s,t,u) {\cal L}_1^{\mu\nu} +B(s,t,u) {\cal L}_2^{\mu\nu}
\end{equation}
is given in terms of the independent Lorentz tensors   
\begin{eqnarray}
\label{invariants}
   {\cal L}_1^{\mu\nu}&=&p_2^\mu p_1^\nu-\frac{1}{2}s g^{\mu\nu}, 
   \nonumber \\ 
   {\cal L}_2^{\mu\nu}&=&-\left(\frac{1}{2}u_1t_1 g^{\mu\nu} 
   +t_1 p_2^\mu p_3^\nu +u_1 p_3^\mu p_1^\nu +s p_3^\mu p_3^\nu\right),
\end{eqnarray}
where we have disregarded the terms that vanish upon the conditions 
$\epsilon_1\cdot p_1=\epsilon_2 \cdot p_2=0$. We have introduced the notation  
$\xi_1=\xi -m_\pi^2$, $\xi=s,t,u$. With the scalar quantities $A$ and $B$ one 
obtains the amplitudes $H_{++}=-(A+m_\pi^2 B)$ and $H_{+-}=(\frac{u_1t_1}{s}-
m_\pi^2)B$ for the equal-helicity and helicity-flipped photons. The 
general expression for the polarizabilities are conventionally obtained in 
the $t$-channel. The superscripts $i=N,C$ denote the neutral and charged 
pions, respectively. The dipole, $\alpha_1^i,\beta_1^i$, and quadrupole, 
$\alpha_2^i,\beta_2^i$, polarizabilities are by definition extracted 
from the first two coefficients of the Taylor expansion of 
the amplitudes around $s=0$ with $u=t=m_\pi^2$ \cite{Guiasu:1979}, 
\begin{eqnarray}
   \frac{\alpha A^i(s,t,u)}{2 m_\pi}&=&\frac{\alpha}{2 m_\pi}\left(
   A^i(0,m_\pi^2,m_\pi^2)+ s \frac{d}{d s} A^i(0,m_\pi^2,m_\pi^2)+\dots\right)
   \nonumber \\ 
   &=&\beta_1^i + \frac{s}{12}\beta_2^i+\dots ,
   \noindent \\
   -\alpha m_\pi B^i(s,t,u)&=&-\alpha m_\pi\left( B^i(0,m_\pi^2,m_\pi^2)
   + s\frac{d}{d s} B^i(0,m_\pi^2,m_\pi^2)+\dots \right )
    \nonumber \\ 
   &=&(\alpha_1+\beta_1)^i + \frac{s}{12}(\alpha_2+ \beta_2)^i+\dots,
\end{eqnarray}
with $\alpha\simeq 1/137$ denoting the QED fine structure constant. The Born 
term, arising in the case of the charged pion, is removed from the amplitudes. 
The NJL Lagrangian used in this work contains pseudoscalar isovector and 
scalar isoscalar four-quark interactions and is minimally coupled to the 
electromagnetic field. Since we do not take into account explicit 
vector meson couplings, the value of the quark weak coupling constant is 
$g_A=1$.\footnote{This affects directly the difference of 
polarizabilities $(\alpha_1-\beta_1)^C$ which is proportional to the 
difference of the low-energy constants ${\overline l}_6-{\overline 
l}_5$ of $\chi$PT, related to the $\pi\to e\nu\gamma$ decay. We stress, 
however, that the expression for $(\alpha_1-\beta_1)^C$ evaluated at the 
leading order of the chiral expansion in the NJL model is formally identical 
to the leading order $\chi$PT result \cite{Bajc:1996}.}

The diagrams of Fig.~\ref{fig:graphs} provide polarizabilities which scale 
as $N_c^0$. Besides the quark one loop diagrams, it is expected that the pion 
loops will have important contributions, mainly in the cases where the 
tree-level results are absent (in the NJL model the chiral counting of meson 
tree-level results are classified in Refs.~\cite{Bajc:1996,Bernard:1992}). 
Although formally the pion-loop contributions to polarizabilities are at a 
suppressed level of $1/N_c$, the small value of the pion mass entering the 
chiral logarithms enhances them. It is out of the scope of the present work 
to calculate the pion-loop contributions self-consistently within the NJL 
model, which is an involved task even for the basic pion quantities 
\cite{Nikolov:1996,Birse:2002}. Instead, we shall include the lowest 
model-independent pion-loop diagram at the $p^4$ order, calculated within 
$\chi$PT in Refs.~\cite{Bijnens:1988,Donoghue:1988,Donoghue:1993}, and known 
to be the only non-vanishing contribution to the amplitude $A$ at this order 
in the neutral channel. The pion-loop amplitudes are
\begin{equation}
\label{mlN} 
   A^N_{\pi l}=-4\frac{s-m_\pi^2}{s f_\pi^2} {\overline G}_\pi(s), 
   \quad A^C_{\pi l}=-2\frac{1}{f_\pi^2}{\overline G}_\pi(s),
\end{equation}
where 
\begin{equation} 
   {\overline G}_\pi(s)=\frac{1}{16 \pi^2}\sum_{n=1}^{\infty}
   \left(\frac{s}{m_\pi^2}\right)^n\frac{(n!)^2}{(n+1)(2n+1)!}.
\end{equation} 
The pion loop in the charged mode contributes only to the quadrupole 
polarizabilities, as it starts out with a term linear in $s$. It has half 
the strength of the neutral quadrupole case. 

The pion loop as well as the $\sigma$-exchange diagram of Fig.~\ref{fig:graphs}
contribute only to the amplitude $A$. Regarding the amplitude $B$ for the 
neutral (charged) mode, only the quark box (quark box + pion exchange) diagrams
contribute in that case, starting from the $p^6$ order for the dipole 
polarizabilities and from the $p^8$ order for the quadrupole polarizabilities. 
Thus the combinations $(\alpha_j+\beta_j)^i$, $(j=1,2)$, to which the $B$ 
amplitude leads, provide a genuine test of the dynamical predictions of the 
NJL model at leading order in $1/N_c$, as they are insensitive to the 
lowest-order $\chi$PT corrections.   

All quark one-loop integrals are regularized using the Pauli-Villars 
prescription with one regulator $\Lambda$ and two subtractions 
\cite{Pauli:1949} (in the context of the NJL model see also 
\cite{Osipov:1985,RuizArriola:1991gc}), 
\begin{equation}
\label{regulator}
   [f(M^2)]_{PV}=f(M^2)-f(M^2+\Lambda^2)+\Lambda^2\frac{\partial 
   f(M^2+\Lambda^2)}{\partial \Lambda^2}.
\end{equation}
This procedure is consistent with the requirements of gauge invariance.
Here $M$ denotes the constituent quark mass, large due to the 
spontaneous breaking of the chiral symmetry. 

It is possible to obtain the results for the quark diagrams in simple 
analytic forms. The expressions are rather lengthy, so we only show a 
typical result,
\begin{equation}
\label{box_L1_dip}
   (\beta^N_1)_{box}=\frac{5\alpha N_cg_\pi^2}{27\pi^2 m_\pi Z_\pi}
   \left[\frac{1}{4 M^2 - m_\pi^2}\left(1+ \frac{m_\pi\arctan\left(
   \frac{m_\pi}{\sqrt{4 M^2 - m_\pi^2}}\right)}{
   \sqrt{4M^2 - m_\pi^2}} \right)\right]_{PV}.
\end{equation}
The factors $g_\pi$ and $Z_\pi$ are the $\pi {\overline q} q$ coupling and 
the pion wave function normalization, respectively 
\cite{Bajc:1996,Bernard:1996}. In the case of the box (box + pion exchange) 
diagrams one is dealing with strictly finite integrals, which remain 
stable upon the removal of the regulator, {\em i.e.} $\lim_{\Lambda\to\infty}
[f(M^2)]_{PV}=f(M^2)$. On the contrary, for the factors $g_\pi$ and $Z_\pi$ 
the regularization is essential.

In Table~\ref{param} we collect the model parameters, obtained by 
fitting the physical pion mass and the weak decay constant. The 
parameters of the model are the four-quark coupling constant $G$, the 
cutoff $\Lambda$, and the current quark mass $m$. These are traded 
for $f_\pi=93.1$~MeV, $m_\pi=139$~MeV (charged mode) or $m_\pi=136$~MeV 
(neutral mode), and the constituent quark mass, $M$.

\begin{table}[tb]
\caption{The NJL model parameters for the charged and neutral channels, with 
the input marked by *. In all cases $f_\pi^*=93.1$~MeV.}
\label{param}
\begin{center}
\begin{tabular}{|c|c|c|c|c|}
\hline
    $M^*$ [MeV] & $m_\pi^*$ [MeV] &$m$ [MeV] & $G$ [GeV${}^{-2}$] 
    &$\Lambda$ [MeV] \\ \hline
250 & 139 & 5.8 & 8.49 & 964 \\
250 & 136 & 5.6 & 8.52 & 963 \\
300 & 139 & 7.5 & 13.1 & 827 \\
300 & 136 & 7.2 & 13.1 & 827 \\
350 & 139 & 8.4 & 17.2 & 767 \\
350 & 136 & 8.1 & 17.3 & 765 \\
\hline
\end{tabular}
\end{center}
\end{table}

Our results are displayed in Tables \ref{tab:N}-\ref{tab:anat}, where we 
show the NJL model predictions for three different values of $M$. 
In Table \ref{tab:N} we collect also the results of the dispersion relations 
(DR) fit to the Crystal Ball data for the process $\gamma\gamma\to\pi^0
\pi^0$ \cite{Marsiske:1990,Bienlein:1992} (for which the dipole 
polarizabilities have been taken from \cite{Fil'kov:1999}); the DSR results
of Fil'kov and Kashevarov \cite{Fil'kov:2005}, and the $\chi$PT predictions 
taken from \cite{Gasser:2005,Bellucci:1994}. In Table \ref{tab:V} we 
include the values of the dipole and quadrupole polarizabilities of the 
charged pion obtained as a result of the DR fit \cite{Fil'kov:2006} to 
the available experimental data for the total cross section in the reation
$\gamma\gamma\to\pi^+\pi^-$ 
\cite{Boyer:1990,Aihara:1986,CELLO:1992,VENUS:1995,Heister:2003,BELLE:2005}
in the energy region from the threshold up to $2.5$ GeV. These results are in 
conformity with the experiments for $\pi^-Z\to\gamma\pi^-Z$ at Serpukhov 
\cite{Antipov:1983,Antipov:1985}, $\gamma p\to\gamma\pi^+n$ at the Lebedev 
Phys. Inst. \cite{Lebedev:1984}, and at MAMI \cite{Ahrens:2005}. The $\chi$PT 
predictions are taken from \cite{Gasser:2006}. 

In Table~\ref{tab:anat} we present the anatomy of our result for the case 
$M=300$~MeV. We display separately the several gauge invariant contributions 
to the polarizabilities: the  box (for neutral polarizabilities), box + pion 
exchange diagram (for the charged polarizabilities), the $\sigma$ exchange, 
and the pion loop (\ref{mlN}). The pion exchange diagram arises only for the 
charged channel and builds together with the box a gauge invariant amplitude.   

\begin{table}[tb]
\caption{The dipole (in units of $10^{-4} {\rm fm}^3$) and quadrupole (in units
of $10^{-4} {\rm fm}^5$) neutral pion polarizabilities. The first three rows 
show our NJL model predictions at various values of the quark mass, $M$.} 
\label{tab:N}
\begin{center}
\begin{tabular}{|l|c|c|c|c|}
\hline
& $(\alpha_1+\beta_1)_{\pi^0}$ & $(\alpha_1-\beta_1)_{\pi^0}$ 
& $(\alpha_2+\beta_2)_{\pi^0}$ & $(\alpha_2-\beta_2)_{\pi^0}$ 
\\ \hline
$M=250$~MeV & 1.13 & -2.05 & -0.33 & 46.3 
\\
$M=300$~MeV & 0.73 & -1.56 & -0.14 & 36.1             
\\
$M=350$~MeV & 0.50 & -1.32 & -0.07 & 30.6 
\\
\hline 
DR fit \cite{Fil'kov:2005} &$0.98\pm 0.03$ &$-1.6\pm 2.2$ 
                               &$-0.181\pm 0.004$ &$39.70\pm 0.02$
\\
DSR \cite{Fil'kov:2005}      &$0.802 \pm 0.035$ &$-3.49\pm 2.13$
                             &$-0.171\pm 0.067$ &$39.72\pm 8.01$ 
\\ 
$\chi$PT \cite{Bellucci:1994,Gasser:2005} 
    &$1.1\pm 0.3$ &$-1.9\pm 0.2$ &$0.037\pm 0.003$ &$37.6 \pm 3.3$ 
\\ 
\hline
\end{tabular}
\end{center}
\end{table}

\begin{table}[tb]
\caption{Same as in Table~\ref{tab:N} for the dipole and quadrupole charged 
pion polarizabilities.} 
\label{tab:V}
\begin{center}
\begin{tabular}{|l|c|c|c|c|}
\hline
          &$(\alpha_1+\beta_1)_{\pi^\pm}$ &$(\alpha_1-\beta_1)_{\pi^\pm}$ 
          &$(\alpha_2+\beta_2)_{\pi^\pm}$ &$(\alpha_2-\beta_2)_{\pi^\pm}$ 
\\ \hline
$M=250$~MeV & 0.28                 & 10.4                 &  0.46                 &   23.8             \\
$M=300$~MeV & 0.19                 & 9.4                  &  0.20                 &   17.5              \\
$M=350$~MeV & 0.13                 & 8.3                  &  0.10                 &   14.2             \\
\hline 
DR fit \cite{Fil'kov:2006} &$0.18^{+0.11}_{-0.02}$ &$13.0^{+2.6}_{-1.9}$  
                                  &$0.133\pm 0.015$     &$25.0^{+0.8}_{-0.3}$
\\ 
DSR \cite{Fil'kov:2006} &$0.166\pm 0.024$ &$13.60\pm 2.15$ 
                        &$0.121\pm 0.064$ &$25.75\pm 7.03$ 
\\ 
$\chi$PT \cite{Bellucci:1994,Gasser:2006} &$0.16$ $[0.16]$ 
                                          &$5.7\pm 1.0$ $[5.5]$ 
                                          &$-0.001$ $[-0.001]$ 
                                          &$16.2$ $[21.6]$ 
\\ 
\hline
\end{tabular}
\end{center}
\end{table}

\begin{table}[tb]
\caption{Contribution of various diagrams to the NJL result for the case 
$M=300$~MeV. Units are the same as in Table \ref{tab:N}.} 
\label{tab:anat}
\begin{center}
\begin{tabular}{|l|c|c|c|c|}
\hline
& box + $\pi$-exchange & $\sigma$-exchange & pion-loop & total  \\ 
\hline  
$(\alpha_1+\beta_1)_{\pi^0}$   &   0.73 &  0    &  0     & 0.73   \\
$(\alpha_1-\beta_1)_{\pi^0}$   & -11.13 & 10.57 & -1.0   & -1.56  \\
$(\alpha_2+\beta_2)_{\pi^0}$   & -0.144 &  0    &  0     & -0.144 \\
$(\alpha_2-\beta_2)_{\pi^0}$   &  5.09  & 9.07  & 21.97  &  36.13 \\ 
\hline
$(\alpha_1+\beta_1)_{\pi^\pm}$ &  0.189 &  0     &   0    & 0.189  \\
$(\alpha_1-\beta_1)_{\pi^\pm}$ & -0.977 & 10.36  &   0    & 9.39   \\
$(\alpha_2+\beta_2)_{\pi^\pm}$ &  0.198 &  0     &   0    & 0.198  \\
$(\alpha_2-\beta_2)_{\pi^\pm}$ & -1.63  &  8.87  &  10.29 &  17.54 \\ 
\hline
\end{tabular}
\end{center}
\end{table}

We observe a dependence on the input choice of the constituent quark mass, 
which affects the polarizabilities more severely in the quadrupole than in 
the dipole case. Let us first consider the sum of the polarizabilities, 
which involves only the box contributions. Comparing the sets with $M=250$ 
MeV and $M=350$ MeV, the box gets reduced in magnitude by about a factor of 
2.2 in the dipole case and by about a factor of 4.6 in the quadrupole case, 
both in the neutral and charged channels. The faster change in the quadrupole 
case can be understood with a crude estimate: the ratio of the leading-order 
term of the dipole to the quadrupole in the expansion around $m_\pi=0$ is 
$\sim M^2$. This gives the suppression factor 
\begin{equation}
   \frac{(\alpha_2+\beta_2)_{M=250~{\rm MeV}}(\alpha_1+\beta_1)_{M=350~{\rm MeV}}}{
   (\alpha_2+\beta_2)_{M=350~{\rm MeV}}(\alpha_1+\beta_1)_{M=250~{\rm MeV}}}
   \sim\frac{M_{M=350~{\rm MeV}}^2}{M_{M=250~{\rm MeV}}^2}\sim 2.  
\end{equation}
The same happens in the charged mode. The data lies half way between the model 
values for $M=250$~MeV and $350$~MeV. The best agreement with respect to the 
DR and DSR data is obtained for the set with $M=300$ MeV, for which we get an 
overall good description. 

We stress that the values (including the overall signs) of the sums 
$(\alpha_2+\beta_2)_{\pi^0,\pi^\pm}$ are totally determined by the 
gauge-invariant quark box or the box + pion exchange contribution. This is a 
straightforward and the most transparent result of the presented NJL-model 
calculation. Moreover, the main part of the box contribution comes from the 
first non-vanishing $p^8$-order term in the chiral expansion.  
Based on this fact we expect that the contact term of the $p^8$ 3-loop 
calculation in $\chi$PT may also play an important role in reversing the 
signs of the 2-loop order results for these quantities.
  
Let us comment on the channels involving the difference of the electric 
and magnetic polarizabilities. i)~$(\alpha_1-\beta_1)_{\pi^0}$: here the box 
contribution is largely canceled by the scalar exchange. At the $p^4$-order 
of the chiral counting they cancel exactly \cite{Bajc:1996}. The higher-order
contributions are quark-mass dependent, decreasing as the constituent quark 
mass increases. The convergence rate is slow, at $p^8$-order one reaches only 
about $50\%$ of the full sum. ii)~$(\alpha_1-\beta_1)_{\pi^\pm}$: contrary to 
the neutral channel, the size of the $\sigma$-exchange diagram for this 
combination is about an order of magnitude larger than the box + pion 
exchange diagram, and it becomes the most important contribution. 
The pion loops are absent. iii)~$(\alpha_2-\beta_2)_{\pi^\pm}$: the pattern 
observed in ii)~repeats itself for the quadrupole polarizabilities. However 
in this case the subleading in the $1/N_c$ counting pion-loop diagram has the 
same magnitude as the $\sigma$-exchange term. 

In conclusion, we highlight our main results: the NJL model at the leading 
order of the $1/N_c$ counting yields the right sign and magnitude for the 
quadrupole polarizabilities of the pion, $(\alpha_2+\beta_2)_{\pi^0,\pi^\pm}$, 
being in conformity with the DR analysis of the data and the DSR predictions. 
The sign is stable when the model parameters are changed. 
The magnitude depends on the value chosen for the unobservable constituent 
quark mass, but the best overall fit to the other empirical data, typically 
yielding $M\sim 300$~MeV, also yields the optimum values for the 
polarizabilities. We have also obtained the convergence rate for several 
combinations of the dipole and quadrupole polarizabilities, comparing the 
$p^8$ chiral order to the full 
result. We have discussed the sensitivity of the different 
polarizability combinations to the inclusion of the one-pion-loop contribution 
that is non-vanishing at the $p^4$ level of the chiral counting. The relative 
contribution of each gauge-invariant combination to the polarizabilities has 
been evaluated and discussed.

\end{document}